\documentclass[opre,nonblindrev]{informs3}

\OneAndAHalfSpacedXI 

\usepackage{enumerate}
\usepackage{algpseudocode}
\usepackage{endnotes}
\usepackage{mathtools}
\let\footnote\endnote

\usepackage{natbib}
\bibpunct[, ]{(}{)}{,}{a}{}{,}%
 \def\bibfont{\small}%
 \def\bibsep{\smallskipamount}%
 \def\bibhang{24pt}%

\usepackage{algorithm}
\usepackage{soul}

\makeatletter
\def\BState{\State\hskip-\ALG@thistlm}
\def\algbackskip{\hskip-\ALG@thistlm}

\TheoremsNumberedThrough     
\ECRepeatTheorems

\EquationsNumberedThrough    

\begin{document}

\RUNAUTHOR{Ashlagi and Saberi and Shameli}

\RUNTITLE{Assignment Mechanisms  under Distributional Constraints}

\TITLE{Assignment Mechanisms  under Distributional Constraints}

\ARTICLEAUTHORS{%
\AUTHOR{Itai Ashlagi}
\AFF{Department of Management Science and Engineering, Stanford University, \EMAIL{iashlagi@stanford.edu}} 
\AUTHOR{Amin Saberi}
\AFF{Department of Management Science and Engineering, Stanford University, \EMAIL{saberi@stanford.edu}} 
\AUTHOR{Ali Shameli}
\AFF{Department of Management Science and Engineering, Stanford University, \EMAIL{shameli@stanford.edu}} 
} 

\ABSTRACT{
We generalize the serial dictatorship (SD) and probabilistic serial  (PS) mechanism for assigning indivisible objects (seats in a school) to agents (students) to accommodate distributional constraints. Such constraints are motivated by equity considerations.
Our generalization of SD maintains several of its desirable properties, including strategyproofness, Pareto optimality, and computational tractability while satisfying the distributional constraints with a small error.

Our generalization of the PS mechanism finds an ordinally efficient and
envy-free assignment while satisfying the distributional constraint with a small error. We show, however,
that no ordinally efficient and envy-free mechanism is also weakly strategyproof. Both of our algorithms
assign at least the same number of students as the optimum fractional assignment.
}

\maketitle

\section{Introduction}
\label{introduction}
We consider the problem of assigning indivisible objects to agents with privately known preferences who are interested in consuming at most one object. One classic solution  for this problem is the serial dictatorship (SD) mechanism, which considers  agents in a certain order and assigns to each agent her most preferred object from the remaining objects. Some applications of SD (with some variations) include school choice \citep{abdulkadirouglu2009strategy,pathak2013school}, college admissions \citep{chen2017chinese,baswana2018centralized},  on-campus housing and office allocation \citep{chen2002improving}.  The  SD mechanism has several desirable properties. It is  strategyproof, Pareto efficient, and  computationally efficient. Random serial dictatorship (RSD), which picks the order of agents uniformly at random, also  treats   agents with identical preferences equally in the sense that it assigns them each object with the same probability.

In spite of RSD's attractive features it  entails an unambiguous efficiency loss ex ante. Specifically, it fails to be ordinally efficient, as agents may be better off trading probability
shares before the outcome is realized (for a formal definition see Section \ref{sec-psm}). The probabilistic serial (PS) mechanism, introduced by \citet{bogomolnaia2001new}, eliminates the inefficiency present in RSD. The PS mechanism  can be described as follows. Imagining that each object is  divisible, all agents simultaneously ``eat" at rate one from their  most preferred object among the remaining objects. This process continues for one unit of time, after which objects are assigned randomly to agents with probabilities that correspond to the  divisible shares each agent has consumed. Note that these mechanisms can be simulated using a computer  after agents report their ordinal preferences over objects. One drawback of PS is that it is not strategyproof, but satisfies a weaker notion, that is known as weak strategyproofness \citep{bogomolnaia2001new}.

The goal of this paper is to extend the SD and PS mechanisms to settings with distributional constraints. Constraints of this sort arise in various contexts. Some school districts impose quotas for students based on geographic location in order to increase socioeconomic integration. Regional quotas are imposed in resident matching in Japan so that programs in  rural areas do not remain underassigned (\citet{kamada2014efficient}). Quotas are  imposed when assigning  cadets to army branches  (\citet{sonmez2013matching}).  Similar policies are adopted in college admissions in various countries \citep{braun2014implementing,biro2010college}. When assigning refugees who  often have family needs, various  constraints arise due to local service capacities \citep{delacretaz2016refugee}. In each of these applications it is desirable to leave as few unassigned agents as possible.

One  challenge with generalizing  these mechanisms  to accommodate  distributional constraints is computational; even checking whether there is an assignment that satisfies the distributional constraints is NP-complete \citep{ashlagi2019assignment} . Given this, we treat the distributional constraints as soft and look for assignments that do not violate the given  constraints by much, as following  \citet{nguyen2016stable}. In some of the above applications it is arguably  reasonable to allow for small violations of these constraints.

Next we describe our contributions.  For exposition purposes, we formulate the problem  in terms of assigning students to schools.  Each student in the model has a publicly known type (a type can encode for example the neighborhood of the student, socio-economic status or race) and a privately known ordinal ranking over schools. Each school has lower- and upper-bound quotas for the number of students of certain types that it can admit. We are interested in assigning as many students to schools as possible. We refer to this property as ``allocative efficiency".

Our main contribution is a  generalization  of the  serial dictatorship mechanism, which maintains strategyproofness, Pareto efficiency, and computational efficiency.  It also produces  assignments that violate the distributional constraints by no more than the number of available types.

In addition to  the above properties, the number of students our mechanism assigns is at least the number of students that can be fractionally assigned, subject to  distributional constraints (see, e.g., \citet{kamada2014efficient} and \citet{ehlers2014school}).  We refer to this benchmark as OPT. Observe  that some constraints may need to be violated  to achieve this benchmark.  One  assumption we make is that all schools are acceptable to all students; this is important  to achieve  allocative efficiency  together with strategyproofness (thus preventing students from truncating their preference lists). This  is a reasonable assumption  when  outside options are very limited, as arguably  the case when assigning refugees, assigning cadets to military schools, or  assigning students with few local private schools.

The key idea behind the mechanism is to carefully design a menu of  schools that are available for each student who is about to be assigned.  This is done by iteratively solving a set of linear programs before each assignment, one for each school, which checks whether the student  can  possibly be assigned to the school in a way that eventually at least  OPT students will  be assigned.

We further introduce a generalization of the probabilistic serial algorithm, which produces an ordinally efficient assignment. The main idea is that during the ``eating" process,  a student who is about to violate a constraint that is associated with the school she is eating from,  switches to eat from her next most preferred school. The eating process terminates with a fractional assignment, which is then  implemented as a lottery over integral assignments such that no distributional constraint is violated by more than the number of existing types.
We further show that the generalized PS is envy-free (within-type) and ordinally efficient. However, in contrast to the setting without distributional constraints, we show that no mechanism is envy-free, ordinally efficient, and weakly strategyproof.

Finally, we note that violating constraints is necessary  for a couple of reasons. First, we wish to assign at least OPT students. Second, the set of constraints is more general than  the bi-hierarchical structure  that is necessary and sufficient for implementing a random assignment using a lottery over feasible assignments \citep{budish2013designing}. In a related work \citet{akbarpour2015approximate} consider a class of more general constraints and show how a given feasible  fractional assignment can be implemented   using lotteries over integral assignments with  small errors (and do not consider the  mechanism design question).

\subsection{Related work}
\label{sec:relatedwork}

There is a growing literature on assignment  and matching mechanisms subject to distributional constraints. Several papers study which constraints allow implementation of affirmative action \citep{kojima2012school,hafalir2013effective,kominers2013designing,westkamp2013analysis,ehlers2014school, braun2014implementing,fleiner2012matroid,yokoi2016generalized,huang2010classified}. These papers consider either lower- or upper-bound constraints. Since satisfying lower-bound constraints is generally impossible \citep{biro2010college}, some of these studies also consider soft constraints, but without providing  guarantees on constraint violations. Nor do these studies consider allocative efficiency.

More general constraints, like ``regional-caps"  have been considered  \citep{kamada2014efficient,kamada2014stability}, and several studies have considered lower- and upper-bound constraints simultaneously \citep{ehlers2014school,fragiadakis2015improving,fragiadakis2016strategyproof,hamada2016hospitals,goto2015improving}. These studies focus on constrained efficiency or weak stability and seek  nonwasteful outcomes.  Our paper allows  for constraints on subsets of types while also seeking allocative efficiency. We focus, however,  only on assignment problems with no priorities.

This paper is inspired by \citet{nguyen:rakesh} and \citet{nguyen2016stable}, which, respectively, study  two-sided markets under complementarities and proportionality constraints   and find mechanisms  that implement stable matchings without violating  each constraint by much. To bound the constraint violations, they  adopt a novel approach using Scarf's Lemma. We build on more straightforward techniques though based as well on linear programming. Moreover, we have a different objective, namely leaving few students unassigned.
\citet{noda2018large} has the same objective as in our paper. However, he assumes constraints cannot be violated and develops, under large market assumptions,  strategyproof mechanisms that are only approximately optimal.

Finally, our paper assumes  schools' preferences are given via  ranking lists and quotas are given exogenously. We refer to \citet{echenique2015control}, who  characterize schools' choice rules that account for diversity preferences and find that  natural axioms yield such quotas.

\section{Model}
\label{sec:model}

A school choice problem consists of a set of  {\bf students} $N=\{1,\ldots,n\}$ and a set of {\bf schools} $M=\{1,\ldots,m\}\cup\{\phi\}$, where $\phi$ is an outside option. Every other school is referred to as a {\bf regular} school.  

Each student $i\in N$ has a {\bf strict preference ordering}  $\succ_i$ over $M$.
We assume that all students prefer every regular school to the outside option and   later we  discuss the robustness of the results based on this assumption.
Each student $i\in N$ is associated with a  commonly known type $t_i$, which belongs to a finite set of types denoted by $T$. Denote by $C_t$ the number of students of type $t \in T$.

An  {\bf assignment} of students to schools is given by a matrix $[(x_{i,s})_{i\in N,s\in M}]$, where $x_{i,s}$ is the probability that $i$ is assigned to $s$,  and for all $i\in N$, $\sum_{s\in M}x_{i,s}=1$.   An assignment  is {\bf integral} if every student $i\in N$ is assigned to a single school $s\in M$ with probability 1. We refer to an integral assignment also as an {\bf allocation}.

It will be useful to  consider the assignment of students  based on their types. A  vector $x=[(x_{t,s})_{t\in T,s\in M}]$  is called a {\bf type-assignment} if for every
type $t\in T$, $\sum_{s\in M}x_{t,s}=C_t$. Note that every assignment corresponds to a unique type-assignment.
Throughout the paper, we will  refer to a type-assignment simply as an assignment.

Next we introduce the distributional constraints. For every   $s\in M$, let $Z(s)\subseteq 2^T$ be a collection of subsets $R\subseteq T$. For every $s\in M$ and every $R\in Z(s)$, we are given lower- and upper-bound {\bf quotas} $\underline{q}_{R, s}$ and  $\bar{q}_{R, s}$, respectively. ($\bar{q}_{T, s}$ can be thought of as the capacity of school $s$.) We assume there are no constraints imposed on $\phi$, i.e. $Z(\phi)=\emptyset$.
Let $\underline{q}=[(\underline{q}_{R,s})_{s\in M, R\in Z(s)}]$ and $\bar{q}=[(\bar{q}_{R,s})_{s\in M, R\in Z(s)}]$. We refer to
 $\textbf{q}=[\underline{q},\bar{q}]$ as the {\bf distributional constraints}.

We say that an assignment $x$ is {\bf feasible}  with respect to $\textbf{q}$   if
$$\underline{q}_{R, s} \le \sum_{t\in R}x_{t, s} \le \bar{q}_{R, s} \quad \forall s\in M, R\in Z(s).$$

An allocation $x$ {\bf Pareto dominates} another allocation $y$ if no student is worse off in $x$ than in $y$ and at least one student is better off.
An allocation $x$ is said to be {\bf Pareto efficient} with respect to $\textbf{q}$ if there is  no other allocation that is feasible with respect to  $\textbf{q}$ that Pareto dominates $x$.

Pareto-efficient allocations can differ with respect to the number of students assigned to regular schools (see Example \ref{ex-pareto} below).
We are interested in maximizing the number of students that are assigned to regular schools. Consider the linear program  (\ref{LP1}), which attains this objective over all feasible fractional assignments, and denote its objective by OPT. That is OPT is the maximum (fractional) number of students that can be assigned to regular schools without violating the distributional constraints.

We say that $x$ is {\bf allocative efficient} if it assigns at least OPT many students to regular schools.
We are interested in finding  allocations that are allocative efficient, while violating each lower- and upper-bound quota by at most $|T|$.

\begin{align*} \label{LP1}\tag{LP1}
\text{OPT}=\max_{x\in \mathbb{R}^{T\times M}} \quad
  &\sum_{t\in T}\sum_{s\in M\setminus\{  \phi \}}x_{t,s}& \nonumber \\
    \text{s.t.} \quad  &\sum_{t\in R} x_{t,s} \leq \bar{q}_{R,s}, &  s\in M, R\in Z(s) \\
    &\sum_{t\in R}x_{t,s} \geq \underline{q}_{R,s}, & s\in M, R\in Z(s) \nonumber \\
    &\sum_{s\in M} x_{t,s}= C_t, & t\in T\nonumber \\
    &x_{t, s} \geq 0, & t \in T, s\in M. \nonumber
\end{align*}
Throughout the paper we assume that (\ref{LP1}) has a feasible solution.

\begin{example}[Few students assigned to regular schools]
\label{ex-pareto}
This example illustrates that Pareto efficiency does not imply allocative efficiency and  a Pareto-efficient assignment can result in many unassigned students. To see this, suppose there  are 3 types of students,  $t_1,t_2,t_3$,  one regular school $s$ with 20 seats. We are given  two constraints: (i) at most 10 students of types $t_1$ or $t_2$ can be assigned to $s$, and (ii) at most 10  students of  types $t_1$ or $t_3$ can be assigned to $s$. Observe that  assigning 10 students of type $t_1$ to $s$, or assigning 10 students of type $t_2$ and 10 students of type $t_3$ to $s$ leads to  Pareto efficiency.
\end{example}

Finally, a \textbf{mechanism} maps  preference profiles to allocations. A mechanism is  \textbf{strategyproof} if it is a weakly dominant strategy for every student  to reveal her true preferences in the game induced by the mechanism. 

\section{Serial Dictatorship with Dynamic Menus}
\label{sec:rsd}

We present  a  generalization of   SD   for assignments with distributional constraints.
The algorithm outputs an assignment that satisfies allocative efficiency and violates every lower- and upper-bound quota by at most $|T|$.
As in SD, students in our algorithm are sequentially  assigned to their most preferred school  from a given menu. A key difference is that the menu given to every student is computed dynamically with the aid of a linear program.

Throughout the  algorithm we maintain a vector $y=[(y_{t, s})_{t\in T,s\in M}]$ that keeps track of the  (possibly fractional) quantity of students of type $t$ assigned to school $s$. We refer to $y$ as an  incomplete assignment. We also  maintain a vector $\Delta=[(\Delta_{t,s})_{t\in T, s\in M}]$ to keep track of how much the lower- and upper-bound quotas corresponding to each type $t\in T$ and school $s\in M$ have changed so far. For the sake of  convenience, define, for any school $s\in M$ and subset of types $R\in Z(s)$, $y_{R, s}=\sum_{t\in R} y_{t, s}$ and $\Delta_{R, s}=\sum_{t\in R} \Delta_{t, s}$.

To design each student's menu, we will need to solve the following  auxiliary linear program (\ref{LP2}). The linear program takes an incomplete assignment $y$ and distributional constraints $[\textbf{\underline{q}}+\Delta, \bar{\textbf{q}}+\Delta]$ as input and looks for a feasible solution $x$ that assigns the remaining students of each type in a way that is both allocative efficient and  feasible with respect to the distributional constraints. The objective of (\ref{LP2}) is to find such a solution that maximizes the quantity of students of a given type $\hat{t}$ that are assigned to a given school $\hat{s}$, which we denote by $f(\hat{t},\hat{s})$.
\begin{align*}\label{LP2}\tag{LP2}
f(\hat{t}, \hat{s})=\max_{x\in \mathbb{R}^{T\times M}} \quad &x_{\hat{t}, \hat{s}} &\\
    \text{s.t.} \quad &\sum_{t\in T}\sum_{s\in M\setminus\{  \phi \}}x_{t, s}+\sum_{s\in M\setminus\{  \phi \}}y_{T, s}  \geq \text{OPT}\\
     &\sum_{t\in R}x_{t, s}+y_{R, s} \leq \bar{q}_{R, s}+\Delta_{R, s},  & s\in M, R\in Z(s)  \\
     &\sum_{t\in R}x_{t, s}+y_{R, s} \geq \underline{q}_{R, s}+\Delta_{R, s},  & s\in M, R\in Z(s)  \\
     &\sum_{s\in M} x_{t, s}+\sum_{s\in M}y_{t, s} = C_t,  &t\in T \\
    &x_{t, s} \geq 0,  &t\in T, s\in M.
\end{align*}



Given the above definitions, we can describe the main ideas of the algorithm. The algorithm considers the students sequentially in a given (or random) order. Each iteration consists of two  steps: (a) assigning the next student to a school (the ``Assignment Step") and (b)  resolving fractional assignments and updating the distributional constraints (the ``Resolution Step").

\vspace{0.4cm}
\noindent\underline{{\bf The Assignment Step.}} Suppose the algorithm has  assigned students  $1,\ldots,i-1$. Let $i$ be the next student to be assigned and suppose $s_i$ is her favorite school. We need to determine whether $i$ can be assigned to $s_i$.
For this, we solve (\ref{LP2}) with $\hat{t}=t_i$ and $\hat{s}=s_i$.  If $f(t_i, s_i)\geq 1$, $i$ is  assigned to $s_i$ (with probability 1). If $f(t_i, s_i) = 0$, $i$ cannot be assigned to $s_i$. In this case the algorithm proceeds to determine if $i$ can be assigned to her next favorite school.

An interesting case arises when $0<f(t_i, s_i)<1$. In this case we must relax some of the distributional constraints to be able to assign student $i$ to $s_i$. Furthermore, prior to observing the preferences of students who are not yet assigned, it is unclear exactly which constraints should be relaxed.  Therefore at this point we  assign only a fraction $f(t_i, s_i)$ of student  $i$ to $s_i$ in the linear program and say that student $i$ is {\bf partially assigned}. Student $i$ remains partially assigned until her remaining fraction is completely assigned to $s_i$ (while the algorithm assigns other students). The assignment is considered {\bf resolved} once she is completely assigned.

\vspace{0.4cm}
\noindent\underline{{\bf The Resolution Step.}} After assigning a student  (either partially or completely), we proceed to the next step, where the algorithm attempts to resolve any existing partial assignments and update the distributional constraints.

To explain how partial assignments are resolved, we use the following definition.  A school $s$ is {\bf critical} for type $t$ if $0<f(t, s)<1$.  Consider an arbitrary  student $j$  that is partially assigned to $s_j$ and let $r_j$ be the remaining fraction of $j$ that is still unassigned. We ask whether there is another school $s\neq s_j$ that is critical for type $t_j$. Namely, we check whether there  exists a school $s\in M$ such that $0<f(t_j, s)<1$. If such a critical school $s$ is found, we update the variables  in a set of operations we label $(j,s)$-{\it updates}:
\begin{equation}
(j,s)\text{-updates:}\quad\quad
\begin{cases}
\label{updates}
\rho\leftarrow \min(f(t_j, s),r_j) \nonumber\\
\Delta_{t_j, s} \leftarrow \Delta_{t_j, s}-\rho \nonumber\\
\Delta_{t_j, s_j} \leftarrow \Delta_{t_j, s_j}+\rho\\
y_{t_j, s_j} \leftarrow y_{t_j, s_j}+\rho \nonumber\\
r_j\leftarrow r_j-\rho
\end{cases}
\end{equation}
The second-to-last operation increases the fraction of  student $j$ assigned to school $s_j$. Note that this ensures that  $j$ will eventually be assigned to $s_j$ with probability 1. The second and third operations adjust the distributional constraints corresponding to schools $s$ and $s_j$ so they are not violated and (\ref{LP2}) remains feasible. More importantly, they ensure  that at every point during the algorithm, there is never more than one partially assigned student of each type.

Observe that for some critical school $s\in M$ and some subset of types $R\in Z(s)$, the lower bound quotas in (\ref{LP2}), i.e., $\underline{q}_{R, s}+\Delta_{R, s}$, may become negative after  a set of $(j,s)$-updates operations. This, however, does not create a problem, since the feasible solution is required to be nonnegative. While the upper bound quotas, $\bar{q}_{R, s}+\Delta_{R, s}$, can also decrease after a set of $(j,s)$-updates operations, they never become negative by the definition of $\rho$.  Also observe that if for some type $t$, there is a less than one unit of students of type $t$ remaining to be assigned, the outside option becomes critical for  $t$. Therefore, while resolving  a partially assigned student, we will treat the outside option as any other school.



Next we provide a formal description of the algorithm, called {\it serial dictatorship with dynamic menus}.
In addition to $y$ and $\Delta$, we also maintain a set $P$ of partially assigned students throughout the algorithm.
\begin{algorithm}[H]
\caption{Serial Dictatorship with Dynamic Menus}
\begin{algorithmic}[1]
\State $\Delta \leftarrow \vec{0}$, $y \leftarrow \vec{0}$, $P \leftarrow \{\}$.
\State \textbf{For} $i=1$ to $n$,
 \State\quad\quad $S\leftarrow M$.
 \State\quad\quad \textbf{While} $i$ is not assigned, \hfill[\textbf{Assignment Step}]
\State\quad\quad\quad\quad $s_i\leftarrow$  $i$'s most preferred school in $S$. \State\quad\quad\quad\quad $S \leftarrow S\setminus s_i$.
\State\quad\quad\quad\quad \textbf{If} $f(t_i, s_i)\geq 1$ then

\State\quad\quad\quad\quad\quad\quad assign $i$ to $s_i$.

\State\quad\quad\quad\quad\quad\quad $y_{t_i, s_i} \leftarrow y_{t_i, s_i}+1$.
\State\quad\quad\quad\quad \textbf{Else if} $ 0 < f(t_i, s_i)\leq 1$ then
\State\quad\quad\quad\quad\quad\quad partially assign $i$ to $s_i$.
\State\quad\quad\quad\quad\quad\quad $P \leftarrow P\cup\{i\}$,  $r_i\leftarrow 1-f(t_i, s_i)$.
\State\quad\quad\quad\quad\quad\quad $y_{t_i, s_i} \leftarrow y_{t_i, s_i}+f(t_i, s_i)$.
\State\quad\quad\textbf{End}
\State\quad\quad\textbf{While} $\exists$ ($j\in P$ and $s\in M\setminus\{s_j\}$) such that $0<f(t_j, s)<1$,  \hfill[\textbf{Resolution Step}]
\State\quad\quad\quad\quad $(j,s)$-updates:
\State\quad\quad\quad\quad $\rho \leftarrow \min(f(t_j,s),r_j)$.
\State\quad\quad\quad\quad$\Delta_{t_j, s} \leftarrow \Delta_{t_j, s}-\rho$.
\State\quad\quad\quad\quad$\Delta_{t_j, s_j} \leftarrow \Delta_{t_j, s_j}+\rho$.
\State\quad\quad\quad\quad$y_{t_j, s_j} \leftarrow y_{t_j, s_j}+\rho$.
\State\quad\quad\quad\quad$r_j\leftarrow r_j-\rho$.
\State\quad\quad\quad\quad\textbf{If} $r_j=0$ then \Comment{Assignment of $j$ to $s_j$ is resolved}
\State\quad\quad\quad\quad\quad\quad$P\leftarrow P\setminus\{j\}$.
\State\quad\quad\textbf{End}
\State\textbf{End}
\end{algorithmic}
\label{alg1}
\end{algorithm}

Our main result is given and analyzed in  Section  \ref{sec:analysis}. In the next section we illustrate  Algorithm \ref{alg1} on a simple example.

\subsection{A simple example}
\label{sec:ex}

To illustrate Algorithm \ref{alg1} consider the following simple example (see Appendix \ref{app-example} for a more involved example).
There are two schools $s_1$ and $s_2$. There are  three students $i$, $j$, and  $k$, whose  types are $t_1$, $t_2$, and $t_3$, respectively. Students $i$ and $j$ prefer $s_1$ over $s_2$ and student $k$ prefers $s_2$ over $s_1$. We are given the following distributional constraints. For each school  $s\in \{s_1, s_2\}$ and every two types $t\neq t'$, $1\leq x_{t, s}+x_{t', s} \leq 2$.



Observe that the unique  feasible fractional solution is the one, in which each student is assigned to each school with probability $0.5$, i.e., $x_{t, s}=0.5$ for every $t\in \{t_1, t_2, t_3\}$ and $s\in\{s_1, s_2\}$.
Assume the order of students to be $i,j$ and $k$. In the first assignment step,  $i$ is partially  assigned to $s_1$ (with 0.5), and is added to the set of partially assigned students. In the first resolution step,  $s_2$ is critical for type $t_1$ and $f(t_1, s_2)=0.5$. Therefore,  $s_2$ is used to resolve the assignment of $i$ to $s_1$ by applying the $(i, s_2)$-updates. After this procedure, the constraints of schools $s_1$ and $s_2$ are updated as follows:
\begin{align*}
	&1.5\leq x_{t_1, s_1}+x_{t_2, s_1} \leq 2.5, &0.5\leq x_{t_1, s_2}+x_{t_2, s_2} \leq 1.5,\\
	&1\leq x_{t_2, s_1}+x_{t_3, s_1} \leq 2, &1\leq x_{t_2, s_2}+x_{t_3, s_2} \leq 2,\\
	&1.5\leq x_{t_3, s_1}+x_{t_1, s_1} \leq 2.5, &0.5\leq x_{t_3, s_2}+x_{t_1, s_2} \leq 1.5.
\end{align*}
In the next assignment step, since $f(t_2, s_1)=0.5$, $j$ will we partially assigned to $s_1$. Similarly, in the next resolution step,  $s_2$ is used to resolve her assignment. The  constraints are updated to:
\begin{align*}
	&2\leq x_{t_1, s_1}+x_{t_2, s_1} \leq 3,&0\leq x_{t_1, s_2}+x_{t_2, s_2} \leq 1,\\
	&1.5\leq x_{t_2, s_1}+x_{t_3, s_1} \leq 2.5,&0.5\leq x_{t_2, s_2}+x_{t_3, s_2} \leq 1.5,\\
	&1.5\leq x_{t_3, s_1}+x_{t_1, s_1} \leq 2.5,&0.5\leq x_{t_3, s_2}+x_{t_1, s_2} \leq 1.5.
\end{align*}
In the final assignment step, $k$ will be partially assigned to $s_2$ since $f(t_3, s_2)=0.5$. Since $f(t_3, s_1)=0.5$, her assignment will be resolved in the following resolution step using the $(k, s_1)$-updates. The final constraints are:
\begin{align*}
	&2\leq x_{t_1, s_1}+x_{t_2, s_1} \leq 3,&0\leq x_{t_1, s_2}+x_{t_2, s_2} \leq 1,\\
	&1\leq x_{t_2, s_1}+x_{t_3, s_1} \leq 2,&1\leq x_{t_2, s_2}+x_{t_3, s_2} \leq 2,\\
	&1\leq x_{t_3, s_1}+x_{t_1, s_1} \leq 2,&1\leq x_{t_3, s_2}+x_{t_1, s_2} \leq 2,
\end{align*}
and the algorithm terminates with $i$ and $j$  assigned to $s_1$ and   $k$  assigned to $s_2$.

\subsection{Analysis of Algorithm \ref{alg1}}
\label{sec:analysis}

Before we prove our main result, we establish a few useful properties of Algorithm \ref{alg1}.

\begin{claim}
\label{lemma0}
Fix a $(j,s)$-updates sequence of operations  for student $j$ and school $s$.
\begin{enumerate}
\item[(i)] Let $x$ be a feasible solution to  (\ref{LP2}) before the $(j,s)$-updates that  satisfies $x_{t_j, s}\geq \rho$. Then  setting $x_{t_j, s}\leftarrow x_{t_j, s}-\rho$  generates a feasible solution to (\ref{LP2}) after the $(j,s)$-updates.
    \item[(ii)] Let $x$  be a feasible solution to (\ref{LP2}) after the $(j,s)$-updates. Then setting $x_{t_j, s}\leftarrow x_{t_j, s}+\rho$  generates a feasible solution to (\ref{LP2}) before the $(j,s)$-updates.
\end{enumerate}
\end{claim}
\proof{Proof.}

We show the first part (the other part follows similar arguments).
Observe that after the $(j, s)$-updates, $\sum_{s\in M\setminus\{\phi\}}y_{T, s}$ increase by $\rho$  and  $\sum_{t\in T}\sum_{s\in M\setminus \{\phi\}} x_{t, s}$ decreases by $\rho$, implying that the  first constraint of (\ref{LP2}) holds.

The second and third constraints in (\ref{LP2})  hold for all  schools other than $s$ and $s_j$ since the values of $x$ and $y$ and $\Delta$ remain the same for these schools. These constraints also hold for school $s$ because $\Delta_{t_j, s}$ and $x_{t_j, s}$ have decreased by the same amount whereas the value of $y_{t_j, s}$ remains unchanged. Similarly they hold for school $s_j$ because $y_{t_j, s_j}$ and $\Delta_{t_j, s_j}$ have increased by the same amount while $x_{t_j, s_j}$ remains unchanged. The fourth constraint holds because after the $(j, s)$-updates, $x_{t_j, s}$ decreases by $\rho$, $y_{t_j, s_j}$ increases by $\rho$, and all other coordinates of $x$ and $y$ remain the same. Therefore, $\sum_{s\in M} x_{t, s}+\sum_{s\in M} y_{t, s}$ remains constant for each $t\in T$. 
\Halmos
\endproof

\begin{lemma}\label{lemma2}
(\ref{LP2}) is feasible after each assignment step and after each resolution step in  Algorithm \ref{alg1}.
\end{lemma}
\proof{Proof.}
At the beginning of the algorithm, before any assignments are made, (\ref{LP2}) is feasible based on the  assumption that there is a feasible solution for (\ref{LP1}). We next show that (\ref{LP2}) is feasible after each assignment step and after each $(j,s)$-update in the resolution step.

Consider an assignment step and let $i$ be the student that is being assigned to school $s_i$ (possibly partially). Observe that the only change in $y$ when $i$ is being assigned is $y_{t_i,s_i}\leftarrow y_{t_i,s_i}+\min(f(t_i,s_i),1)$.  Let $x$ be a solution for $f(t_i,s_i)$ just before assigning $i$ to $s_i$.  Then setting $x_{t_i, s_i}\leftarrow x_{t_i, s_i}-\min(x_{t_i, s_i}, 1)$ while keeping all other coordinates of $x$ the same generates a feasible solution to (\ref{LP2}) immediately after $i$ is assigned.

Now consider a resolution step and assume that  $x$ is a feasible solution of $f(t_j, s)$ just before the $(j,s)$-updates step and let $r$ be the remaining unassigned fraction of student $j$. Then, by Claim \ref{lemma0},  setting $x_{t_j, s}\leftarrow x_{t_j, s}-\min(x_{t_j, s}, r)$ while keeping all other coordinates of $x$ the same generates a feasible solution to (\ref{LP2}) after the $(j,s)$-updates. This completes the proof.
\Halmos
\endproof

\begin{lemma}\label{lemma1}
After Algorithm \ref{alg1} terminates, no student remains partially assigned.
\end{lemma}
\proof{Proof.}
We first show that at any point during the running of algorithm, at most one student per type is partially assigned. For the sake of contradiction, suppose students $i$ and $j$ ($i > j$) are the first two students of the same type that are both partially assigned at some point in the algorithm. 




Consider the resolution step just before the algorithm proceeds to assign student $i$. By design, while the assignment of student $j$ is not resolved and there exists a critical school $s'$ for type $t_j$, the algorithm uses that school to resolve the assignment of $j$. Since by the end of the resolution step, the assignment of student $j$ is not resolved, there must be no remaining critical schools for type $t_j$. Therefore student $i$ cannot be partially assigned.


We can now show that no student is partially assigned after the algorithm terminates. Suppose, for the sake of contradiction, that this is not the case and there is a student $j$ of type $t$ that  remains partially assigned and let $r_j$ be the  fraction of $j$ that  remains. By the above argument, $j$ is the only student of type $t$ that is partially assigned. We claim that in the resolution step of the last iteration, assignment of $j$ to $s_j$ will be resolved. Suppose this is not the case, then by Lemma \ref{lemma2}, (\ref{LP2}) is still feasible after termination and has some solution $x$. Since $j$ is the only student of type $t$ that is partially assigned, it must be that  $\sum_{s\in M} x_{t, s}=r_j$. Therefore, since $r_j<1$, this implies that there exists some school that is critical with respect to $t$, contradicting the assumption that the algorithm has terminated.

\Halmos
\endproof

\begin{lemma}\label{lemma3}
After Algorithm \ref{alg1} terminates, the only feasible solution for (\ref{LP2}) is $\vec{0}$.
\end{lemma}
\proof{Proof.}
Assume the algorithm has terminated. By Lemma \ref{lemma2}, (\ref{LP2}) is  feasible and by Lemma \ref{lemma1}, all students are  assigned to some school (including, possibly, to the outside option) with probability 1. This implies that for every $t\in T$, $\sum_{s\in M} y_{t, s}=C_t$. Thus $\vec{0}$ is the unique feasible solution for (\ref{LP2}).
\Halmos
\endproof

\begin{lemma}\label{lemma4}
For every $s'\in M$ and every $t'\in T$, $f(t',s')$ does not increase after each assignment step and each resolution step in  Algorithm \ref{alg1}.
\end{lemma}
\proof{Proof.}
Let $s'\in M$ be an arbitrary school and $t'\in T$  an arbitrary  type.
After an assignment step, $f(t', s')$ cannot increase since $y$ only increases after an assignment. We next show that during the resolution step, $f(t', s')$  cannot increase after updating $\Delta$ and $y$ (i.e., after any $(j,s)$-updates). Consider a $(j,s)$-updates step for some student $j$ and a critical school $s$ for $t_j$.
Let $P_1$ be the set of all feasible solutions for (\ref{LP2}) right before the $(j,s)$-updates and $P_2$ be the set of all feasible solutions for (\ref{LP2}) after the $(j,s)$-updates.
Fix a feasible solution  $x\in P_2$. By Claim \ref{lemma0}, updating $x$ so that $x_{t_j, s}\leftarrow x_{t_j, s}+\rho$, where $\rho$ is defined as in the $(j,s)$-updates, generates a feasible solution in $P_1$.
Since the value of $f(t',s')$ before and after the $(j, s)$-updates is defined as the maximum of $x_{t',s'}$ over all feasible solutions $x$ in $P_1$ and $P_2$ respectively, it cannot be increasing.
\Halmos
\endproof

\begin{lemma}\label{lemma5}
After Algorithm \ref{alg1} terminates, $|\Delta_{R,s}|\leq |T|$ for every $s\in M$ and every $R\in Z(s)$.
\end{lemma}
\proof{Proof.}
Since for any $R\in Z(s)$, $\Delta_{R, s}=\sum_{t\in R} \Delta_{t, s}$, it suffices to show that for any school $s$ and for any type $t$, $-1 \leq \Delta_{t, s} \leq 1$.
We show in fact  that these inequalities hold at any time during the algorithm.

Let $j$ be a student that is partially assigned to $s_j$. We first argue that if $\Delta_{t_j, s_j}$ increases (line 19), it must be true that $f(t_j,s_j)=0$. To see this, note that after  $j$ is partially assigned to $s_j$ and $y_{t_j,s_j}$ is increased (line 13), $f(t_j,s_j)=0$, and therefore, by Lemma \ref{lemma4}, $f(t_j,s_j)$  remains zero thereafter. Moreover, when $s$ is critical for $t_j$, $\Delta_{t_j, s}$ can  only decrease. This means that for all $s'\in M$ and $t'\in T$, as long as $f(t',s')>0$, $\Delta_{t', s'}$ can only decrease. And once $f(t',s')$ becomes zero (which happens by Lemma $\ref{lemma3}$),  $\Delta_{t', s'}$ only increases.

Suppose $\Delta_{t_j, s}$ decreases by some amount $\rho$ (at line 18) and suppose $f(t_j, s)=a$  immediately  before this decrease. By Claim \ref{lemma0},  $f(t_j,s)=a-\rho$ after the $(j,s)$-updates.
Therefore, since school $s$ is critical with respect to $t_j$, $\Delta_{t_j, s}$ can decrease by at most $1$. On the other hand,  $\Delta_{t_j, s_j}$ increases when we attempt to resolve the partial assignment of  student $j$ to school $s_j$. Let $r_j$ be the remaining fraction of $j$ to be assigned. Whenever $\Delta_{t_j, s_j}$  increases by $\rho$ (line 19), $r_j$ decreases by $\rho$ (line 21). Therefore, by definition, $\Delta_{t_j, s_j}$ cannot increase by more than 1.
\Halmos
\endproof

\begin{theorem}\label{th1}
Consider a school choice problem with distributional constraints $\textbf{q}$. Algorithm \ref{alg1} outputs an allocation $y$ and a vector $\Delta$ such that:
\begin{enumerate}
\item[(i)] $y$ is feasible with respect to $[\underline{q}+\Delta, \bar{q}+\Delta]$, \label{one}
\item[(ii)] $y$ is allocative efficient,
\item[(iii)] $|\Delta_{R,s}|\leq |T|$ for every $s\in M$ and every $R\in Z(s)$, and 
\item[(iv)] $y$ is Pareto efficient with respect to $[\underline{q}+\Delta, \bar{q}+\Delta]$.
\end{enumerate}
Moreover, the mechanism induced by  Algorithm 1 is strategyproof.
\end{theorem}
\proof{Proof.}
The first three properties follow directly from Lemmas \ref{lemma2}, \ref{lemma3}, and \ref{lemma5}.

Next we show that the mechanism induced by Algorithm \ref{alg1} is strategyproof. Fix some arbitrary student $i$. We claim that $i$'s preferences cannot affect the assignments of all students that are assigned prior to $i\in N$. Note that once a student is partially assigned to a school, she will eventually be assigned to that school with probability 1 by Lemma \ref{lemma1} and since the remainder of a partially assigned student $j$ is always assigned to $s_j$. So the only way in which $i$ can affect the assignment of a student $j<i$ is through her type, which cannot be altered. Finally, when it is $i$'s turn to be assigned, she has no reason to misreport her preferences since, again,  even if she is partially assigned to $s_i$, she will be eventually assigned to that school with probability 1.

We proceed to prove part (iv). For this we need  the following two claims.

\begin{claim}\label{claim1}
	Throughout the algorithm, for any $t\in T$ and $s\in M$, $y_{t, s}-\Delta_{t, s}$ does not decrease.
\end{claim}
\proof{Proof.}
Observe that during  the assignment step the claim holds since $y$ can only  increase and $\Delta$ does not change.
Consider a $(j,s)$-updates step. Either the value of $\Delta_{t_j,s}$ decreases (line 18) while the corresponding value  $y_{t_j,s}$ remains unchanged, or the values of $\Delta_{t_j,s_j}$ and $y_{t_j,s_j}$ increase together by the same amount (lines 19-20).
\Halmos
\endproof

\begin{claim}\label{claim-bijection}
Consider the values $y$ and $\Delta$ after a resolution step.
\begin{enumerate}

\item[(i)] Suppose $x$ is a  feasible solution for (\ref{LP2}). Then $x+y-\Delta$ is an optimal feasible solution for (\ref{LP1}).
\item[(ii)] Let $x$ be an  optimal feasible solution for (\ref{LP1}) such that $x-y+\Delta\geq \vec{0}$. Then  $x-y+\Delta$ is a feasible solution for (\ref{LP2}).
\end{enumerate}
\end{claim}
\proof{Proof.}
We prove the first part (the second part follows from similar arguments). Note, by Lemma \ref{lemma2}, that the set of feasible solutions  for (\ref{LP2})  is not empty. 

Observe (from the  updates in the resolution step) that  $\sum_{s\in M} \Delta_{t, s}=0$ for all $t\in T$. This implies that  if $x$ is feasible for (\ref{LP2}), then $x+y-\Delta$ assigns at least OPT students to regular schools because

$$\sum_{t\in T} \sum_{s\in M\setminus\{  \phi \}} (x+y-\Delta)_{t, s}=\sum_{t\in T} \sum_{s\in M\setminus\{  \phi \}} x_{t, s}+\sum_{s\in M\setminus\{  \phi \}}y_{T, s} \geq \text{OPT}.$$

Since $x$ is a feasible solution for  (\ref{LP2}), for every $s\in M$ and every $R\in Z(s)$,

$$\sum_{t\in R} (x+y)_{t, s} \leq \bar{q}_{R,s}+\Delta_{R, s}, $$
and therefore the first constraint in (\ref{LP1}) also holds, namely for every $s\in M$ and every $R\in Z(s)$,
$$\sum_{t\in R} (x+y-\Delta)_{t, s} \leq \bar{q}_{R,s}.$$

Other constraints can be similarly verified. Moreover, by Claim \ref{claim1} and because $y\equiv \vec{0}$ and $\Delta\equiv \vec{0}$ at the beginning of the algorithm, it must be true that $(x+y-\Delta)_{t, s}\geq 0$ for all $t\in T$ and $s\in M$.   
\Halmos
\endproof

We can now complete the proof. Let $y$ and $\Delta$ be the outcomes of the algorithm (as in the statement of  Theorem \ref{th1}). For the sake of contradiction, suppose there exists an allocation $y'$ that is feasible with respect to $[\underline{q}+\Delta, \bar{q}+\Delta]$ and Pareto dominates $y$. Let student $i$ be the first student (with respect to the order of the algorithm) who is assigned to different schools under $y$ and $y'$, and let these schools be $s$ and $s'$, respectively. Note that $i$ prefers $s'$ to $s$.

Let  $\gamma$ be the time during which the algorithm reaches the assignment step at iteration $i$.
Let $y_{\gamma}$ and $\Delta_{\gamma}$ be the values of $y$ and $\Delta$ at  time $\gamma$.

Since the algorithm did not assign $i$ to $s'$, it must be true that $f(t_i, s')=0$. Therefore, by Claim \ref{claim-bijection},  at time $\gamma$ there is no optimal feasible solution $x$  for  (\ref{LP1}) such that  $x-y_{\gamma}+\Delta_{\gamma}\geq \vec{0}$ and  $(x-y_{\gamma}+\Delta_{\gamma})_{t_i, s'}>0$. We will obtain a contradiction by showing that such an $x$ exists.

We show that, upon termination, there  exists  $0<c\leq 1$, such that (i) $y-\Delta+c(y'-y)$ is an optimal feasible solution for (\ref{LP1}),  (ii) $y-\Delta+c(y'-y)-y_{\gamma}+\Delta_{\gamma}\geq \vec{0}$, and (iii) $[y-\Delta+c(y'-y)-y_{\gamma}+\Delta_{\gamma}]_{t_i,s'}> 0$. This will complete the proof.

First we show that (i) holds for any $0<c\leq 1$. It is equivalent to show that there exists a $c$ such that $0<c\leq 1$ and that $(1-c)(y-\Delta)+c(y'-\Delta)$ is an optimal feasible solution for (\ref{LP1}). By Lemma \ref{lemma3}, after termination, $\vec{0}$ is  feasible for (\ref{LP2}). Therefore, by Claim \ref{claim-bijection}, $y-\Delta$ is an optimal feasible solution for (\ref{LP1}). Consider (\ref{LP2}) upon termination. Since $y'$ is feasible with respect to $[\underline{q}+\Delta, \bar{q}+\Delta]$ and assigns at least OPT students (since it Pareto dominates $y$), $\vec{0}$ is feasible for (\ref{LP2}) when $y$ is replaced with $y'$. 
Hence, by a similar argument as in Claim \ref{claim-bijection}, $y'-\Delta$ is an optimal feasible solution for (\ref{LP1}). This implies (i) because for any $0<c\leq 1$,  $(1-c)(y-\Delta)+c(y'-\Delta)$ is a convex combination of optimal feasible solutions.

Next we show that (ii) holds. By Claim \ref{claim1}, $y-\Delta\geq y_{\gamma}-\Delta_{\gamma}$. If $y'\geq y$, then (ii) holds.  Suppose this is not the case and let $s\in M$ and $t\in T$ be such that  $(y'-y)_{t,s}<0$. It is sufficient to show that for such $s$ and $t$,   $(y-\Delta)_{t,s}> (y_{\gamma}-\Delta_{\gamma})_{t,s}$.
Since $y'\geq y_{\gamma}$ and $y_{t,s}>y'_{t,s}$,  we have that $y_{t, s}>(y_{\gamma})_{t, s}$.
Moreover, observe that when a student of type $t$ is assigned to school $s$ (possibly partially),  $(y-\Delta)_{t, s}$  strictly increases by $\min(f(t, s), 1)$.
Therefore, $(y-\Delta)_{t, s}>(y_{\gamma}-\Delta_{\gamma})_{t, s}$, which completes the proof.

 Finally, we show that (iii) holds. By Claim \ref{claim1}, $y-\Delta \geq y_{\gamma}-\Delta_{\gamma}$. Recall that at time $\gamma$, $f(t_i, s')=0$. Therefore, by Lemma \ref{lemma4}, after time $\gamma$ the algorithm  does not assign any other student of type $t_i$ to $s'$, implying that $y_{t_i, s'}=(y_{\gamma})_{t_i, s'}$. Since every student $j<i$  is assigned the same school under $y$ and $y'$ and $i$ is assigned to $s'$ under $y'$, it  must be true that $y'_{t_i, s'}>y_{t_i, s'}$. This implies that (iii) holds for any $0<c\leq 1$.
\Halmos
\endproof

\noindent{\bf Remarks:}
\begin{enumerate}

\item Algorithm \ref{alg1}  runs in polynomial time. To see this, note that in each assignment step, we solve (\ref{LP2}) at most $|M|$ times. Also, by Claim \ref{lemma0}, in each $(j, s)$-update either the assignment of student $j$ gets resolved or $f(t_j, s)$ becomes zero, in which case, by Lemma \ref{lemma4}, it  remains zero forever. Therefore in each resolution step, for each school $s\in M$ and each student $j\in N$, the  $(j, s)$-updates are done at most once.


\item When Algorithm \ref{alg1} selects the order in which students are assigned uniformly at random, the outcome is symmetric. That is, any two students with the same type and identical preferences have the same probabilistic assignment. 

\item We have assumed that all students prefer every regular school to the outside option $\phi$ (and in particular all regular schools are acceptable). The results and analysis carry through, however, if the set of acceptable schools for  each student is publicly known (which may be a reasonable assumption in the  military or refugee assignment problems).
    This assumption is necessary  to satisfy the lower bounds and allocative efficiency as well as to maintain strategyproofness. Indeed, if students can submit a partial preference list (e.g., by truncating their preferences), strategyproofness may fail to hold.

\item When the set of constraints are laminar and all  lower- and upper-bound quotas are integers, Algorithm \ref{alg1} finds an allocation that is feasible with respect to all the constraints without the need to violate any of the lower- and upper-bound quotas. We formalize this beginning with defining laminar constraints. For every $s\in M$, we say that $Z(s)$ is {\bf laminar} if for each $R,R'\in Z(s)$, such that $R\cap R'\neq \emptyset$,  either $R\subseteq R'$ or $R'\subseteq R$.  We call an assignment problem  {\it laminar} if $Z(s)$ is laminar for all $s\in M$.


\begin{proposition}
\label{prop-laminar}
Given a laminar assignment problem, Algorithm \ref{alg1} finds an integral assignment that is feasible with respect to {\bf q}.
\end{proposition}

Before we prove the proposition, we explain how a laminar assignment problem  can be reduced to a max flow problem with integer lower and upper bound capacities on  edges. This will imply the  existence  of an integral and optimal solution of (\ref{LP1}) that  is feasible with respect to {\bf q}, which can also be found in polynomial time.

A few preparations are useful to illustrate this reduction.
For every $s\in M\setminus\{\phi\}$ and every $R\in Z(s)$, let
\[X(R,s)=\{R'\in Z(s) |\,\,R'\subset R \,\,\,\text{and}\,\,\,(\nexists R''\in Z(s):R'\subset R''\subset R)\}.\]

We  now explain how to construct the graph that corresponds to the  max flow problem. For every $s\in M\setminus\{\phi\}$ and every $R\in Z(s)$, add a node $u_{R, s}$. For every $s$ and every $R\in Z(s)$ and every $R'\in X(R,s)$,  add a directed edge from $u_{R', s}$ to $u_{R, s}$, with lower and upper flow constraints  $\underline{q}_{R', s}$ and $\bar{q}_{R', s}$ respectively.

We  add a source $A$ and a sink $B$. From each node $u_{R,s}$ (not including the source or the sink), with no outgoing edges, add an edge to $B$ with lower and upper bound flow constraints $\underline{q}_{R, s}$ and $\bar{q}_{R, s}$ respectively. We add  $|T|$ {\it auxiliary} nodes $s_1,s_2,\ldots, s_{|T|}$ each of which represents a type and add an edge from $A$ to each $s_t$ with capacity $C_t$.

Finally, for each type $t$ and every school $s\in M\setminus \{\phi\}$, add an edge from $s_t$ to $u_{R, s}$ with infinite capacity for every $R\in Z(s)$ such that $t\in R$ and $u_{R, s}$ has no incoming edges.

Now, observe that every feasible solution to the laminar assignment problem, i.e. (\ref{LP1}) corresponds to a feasible flow in the constructed graph with same objective value and vice versa.

\proof{Proof of Proposition \ref{prop-laminar}.}

Since the laminar assignment problem can be reduced to a max flow problem with integral flow constraints on the edges, the polytope corresponding to the set of optimal solutions for (\ref{LP1}) has integral extreme points. We show that this also holds for (\ref{LP2}) at any point during the algorithm.

First we argue that this holds if no student is ever partially assigned. To see this, note that after each student is assigned, all the lower- and upper-bound quotas change by an integral amount. Therefore, similar to the above, one can reduce our problem  into a max flow problem with integral flow constraints on the edges which implies that the polytope corresponding to the set of feasible solutions for (\ref{LP2}) has integral extreme points.

It remains to show that  Algorithm \ref{alg1} never partially assigns any student. Suppose this is not the case and let $i$ be the first student that is partially assigned to some school $s_i$. Just before assigning $i$, let $\mathcal{P}$ be the set of feasible solutions for (\ref{LP2}). Observe  that by the above argument, all extreme points of $\mathcal{P}$ are integral. Since $i$ was partially assigned to $s_i$, this implies that
$$0<\max_{x\in \mathcal{P}} x_{t_i, s}<1.$$
This is a contradiction, because every point in $\mathcal{P}$ can be written as a convex combination of extreme points.

This means that there must exist an integral point $x'\in \mathcal{P}$ with $x'_{t_i, s}=1$.
\Halmos
\endproof

\end{enumerate}

\section{Generalized Probabilistic Serial Mechanism}
\label{sec-psm}

In this section we generalize the probabilistic serial mechanism (PS)  to allow for distributional constraints. The PS mechanism was introduced by
\citet{bogomolnaia2001new}, who showed that it satisfies several desirable properties such as ordinal efficiency, envy-freeness, and weak strategyproofness.

Let us begin with a brief description of PS, also known as the {\it eating} algorithm. Treating each school (including the outside option) as a divisible object, the algorithm asks the students to eat from schools simultaneously and at the same rate until each consumes one unit. Every student begins eating from her favorite school.
Whenever a school $s$ is fully consumed, students who were eating from $s$ then proceed to eat from their next preferred available school. The process concludes when each student consumes one unit. The resulting fractional outcome is implemented using a lottery over allocations defined by the Birkhoff-Von-Neumann Theorem (\citet{schrijver2003combinatorial}).

To  define the  properties of PS, we  use the (student) assignment variables $x_{i,s}$, which is interpreted as  the probability that student $i$ is assigned to school $s$.
For  assignment $x$, we denote by $x_i=(x_{i,s})_{s\in S}$ the assignment for student $i$, which is the distributional outcome for student $i$.


Let $x$ and $y$ be assignments. We say that $x_i$ \textbf{stochastically dominates} $y_i$ with respect to preference order $\succ_i$, if for every $s\in M$:
$$\sum_{s':s' \succ_i s} x_{i, s'} \geq \sum_{s':s' \succ_i s} y_{i,s'},$$ in which case we  write $x_i sd(\succ_i) y_i$.

Given a preference profile $(\succ_i)_{i\in N}$, we say that $y$ is {\bf stochastically dominated} by $x$ if  $x_i sd(\succ_i) y_i$ for all  $i\in N$ and $x\neq y$.
The assignment $x$ is said to be \textbf{ordinally efficient}, if it is a feasible solution (for (\ref{LP3})) and it is not stochastically dominated by any other feasible assignment.

The assignment $x$ is \textbf{within-type envy-free} if for any two students  $i, j$ of the same type $t$ and every school $s\in M$:
$$\sum_{s': s'\succ_i s} x_{i, s'} \geq \sum_{s':s'\succ_{i} s} x_{j, s'}.$$

Finally, a mechanism  is \textbf{weakly strategyproof} if for every student  $i$ and any preference profile of all other students, reporting $\succ_i$ results in an assignment $x_i$ for $i$, that is not stochastically dominated with respect to $\succ_i$ by any other assignment $x'_i$ for $i$ that can be obtained by $i$ misreporting her preferences, unless $x_i=x'_i$.


Our algorithm, which we call the  \emph{generalized probabilistic serial (GPS)}, generalizes the PS algorithm to the setting with distributional constraints. The outcome of the GPS is a fractional assignment that does not violate any lower- or upper-bound quotas. We show that such a  fractional assignment can be implemented as a lottery over integral solutions that violate each quota by at most $|T|$.


We  begin with establishing the  implementation of such a  fractional solution as a distribution over allocations and explain the details of the GPS algorithm in Section \ref{eating}.

\subsection{Implementing a fractional solution as a lottery over allocations}

Recall the linear program (\ref{LP1}) for optimizing allocative efficiency given the distributional constraints. This program can be rewritten using the (student) assignment variables $x_{i,s}$, which can be interpreted as  the probability that student $i$ is assigned to school $s$:
\begin{align*}\label{LP3}\tag{LP3}
\text{OPT} =\max_{x\in \mathbb{R}^{N\times M}} \quad
    & \sum_{i\in N}\sum_{s\in M\setminus\{  \phi \}} x_{i, s} & \nonumber \\
    \text{s.t.} \quad
    &\sum_{i\in N: t_i \in R} x_{i, s} \leq \bar{q}_{R, s}, &\forall s\in M, R\in Z(s) \nonumber \\
    &\sum_{i\in N: t_i \in R} x_{i, s} \geq \underline{q}_{R, s}, &\forall s\in M, R\in Z(s) \\
    &\sum_{\substack{s\in M}} x_{i, s} = 1, &\forall i\in N\nonumber \\
    &x_{i, s} \geq 0, &\forall i \in N, s\in M. \nonumber
\end{align*}

We  will show that every optimal solution for (\ref{LP3}) can be written as a convex combination of approximately feasible allocations.

\begin{definition}
    An allocation $x$ is {\bf approximately feasible} if it assigns students in a way that
    \begin{enumerate}
        \item Each lower- and upper-bound quota is violated by at most $|T|$. That is
    $$\underline{q}_{R, s} - |T| \le \sum_{t\in R}  x_{t, s} \le \bar{q}_{R, s} + |T| \quad \forall s\in M, R\in Z(s).$$
        \item At least $\lfloor \text{OPT} \rfloor$ students are assigned to regular schools.
    \end{enumerate}
\end{definition}

%

\begin{lemma}\label{implementtheorem}
Every optimal solution for (\ref{LP3}) can be written as a convex combination of approximately feasible integral solutions.
\end{lemma}
\proof{Proof.}
Let $x$ be an optimal solution for (\ref{LP3}). Consider the following polyhedron with variables $y$:
\begin{align*}
    &\Big\lfloor\sum_{i\in N: t_i = t} x_{i, s}\Big\rfloor \leq \sum_{i\in N: t_i = t} y_{i, s} \leq \Big\lceil\sum_{i\in N: t_i = t} x_{i, s}\Big\rceil, &\forall t\in T, s\in M\setminus\{\phi\} \\
    &\Big\lfloor\sum_{i\in N} x_{i, \phi}\Big\rfloor \leq \sum_{i\in N} y_{i, \phi} \leq \Big\lceil\sum_{i\in N} x_{i, \phi}\Big\rceil &\\
    &\sum_{\substack{s\in M}} y_{i, s} = 1, &\forall i\in N\nonumber \\
    &y_{i, s} \geq 0, &\forall i \in N, s\in M. \nonumber
\end{align*}
Note that $x$ is a feasible solution for the above linear program, so the set of feasible solutions is non-empty. Second, the above linear program corresponds to a generalized assignment problem with integer lower- and upper-bound capacities and therefore the corner points of its corresponding polytope are integral.

Finally, every corner point $y$ of the above polytope is approximately feasible. The reason is that  for each upper-bound quota $\bar{q}_{R, s}$ over $s\in M\setminus \{\phi\}$ and $R\in Z(s)$, we have
$$\sum_{i\in N:t_i\in R} y_{i, s}\leq \sum_{t\in R}\Big\lceil\sum_{i\in N:t_i=t} x_{i, s}\Big\rceil\leq \sum_{t\in R}\Big(\sum_{i\in N:t_i=t} x_{i, s}+1\Big)\leq \sum_{i\in N:t_i\in R} x_{i, s}+|T|\leq \bar{q}_{R, s}+|T|.$$

A similar argument can be used for lower-bound quotas. Furthermore, the number of students who are  assigned to the outside option by $y$ is:

$$\sum_{i\in N} y_{i, \phi}\leq \Big\lceil\sum_{i\in N} x_{i, \phi}\Big\rceil.$$
Therefore at least $\lfloor \text{OPT} \rfloor$  students are assigned to regular schools.
\Halmos
\endproof

We note that   Lemma \ref{implementtheorem} could also be derived from \citep{akbarpour2015approximate}. Also, as remarked in the previous section, if the set of constraints are bi-hierarchical,  one can implement the assignment without violating any lower- or upper-bound quota.

\subsection{A generalization of the probabilistic serial mechanism}
\label{eating}

The eating algorithm, PS, is very similar to  \citet{bogomolnaia2001new} in that every student eats from her favorite school as long the ``partial" assignment  is ``extendable'' to an optimal solution for (\ref{LP3}) and switches to her next favorite  school when a constraint becomes tight.

\begin{definition}\label{extendable}
A vector $y$ is {\bf extendable} if there exists $x$, a feasible solution for (\ref{LP3}) that also satisfies the following conditions:
\begin{enumerate}[(i)]
	\item $x$ dominates $y$, i.e., $x_{i,s} \geq y_{i,s}$ for all  $i\in N$ and $s\in M\setminus\{\phi\}$, and
	\item $x$ is allocative efficient, i.e., $\sum_{i\in N} \sum_{s\in M\setminus\{  \phi \}} x_{i, s}=\text{OPT}$.
\end{enumerate}

\end{definition}

Given any vector $y$, it is possible to check whether it is extendable by adding the  linear constraints corresponding to conditions (i) and (ii) to (\ref{LP3}) and testing whether the set of feasible solutions remains non-empty.  Let $\mathcal{E}$ denote the set of extendable vectors. 

Similar to  PS, the GPS starts from an empty assignment and asks every student to eat from the schools at a constant rate in their order of preference. The main difference, however, is that we keep the assignment vector in $\mathcal{E}$ at all times. Whenever we reach a boundary  at which we are about to leave the set $\mathcal{E}$, we prevent the corresponding students from consuming their current school and ask them to move to their next most preferred available school. We then continue the  process  and have all agents consume their current preferred and available school. We do this until the algorithm terminates, i.e., every  student has consumed one unit.

The above process can be implemented as follows. Consider any point during  the running of the algorithm and let $x_{i, s}$ denote how much of school $s$ is consumed by student $i$ at that time. Also consider $\theta$, which denotes the eating pattern of the students at this time, i.e., $\theta_{i,s} = 1$ if $i$ is currently eating from $s$ and $0$ otherwise. Using a linear program, we can find the maximum $c$ such that $x + c \theta$ is extendable. This will determine how long we can continue the current eating pattern.
Whenever we reach that point, a new constraint will be tight. At that point, we ask all the students involved in that constraint to stop eating their current school and move to their next option. We repeat this procedure until all students have consumed one unit in which point $x$ is a (fractional) optimal solution for (\ref{LP3}).

The running time of this process is polynomial, because every student switches at most $|M|$ times. Once a student reaches the outside option, she can continue consuming that option until her total consumption reaches one unit.


\begin{theorem}
The (fractional) assignment produced by the generalized probabilistic serial (GPS) algorithm is
\begin{enumerate}[(i)]
	\item within-type envy-free,
	\item ordinally efficient, and
\item implementable using a lottery over approximately feasible assignments.
\end{enumerate}
\end{theorem}

\proof{Proof.}
The following claim will be  needed:
\begin{claim}\label{claim3}
	Once a student of type $t$ is prevented from eating from some school $s$, then no students of type $t$ will ever be able to eat from that school again.
\end{claim}
\proof{Proof.}
	Consider the first time the algorithm starts blocking students of type $t$ from eating school $s$. Let $y$ denote the set of probability shares consumed by students up to that point such that $y_{i, s}$ represents how much of school $s$ is consumed by student $i$. Since the algorithm treats all students of the same type the same way, at that point all the students of type $t$ are blocked from consuming $s$. By Definition \ref{extendable}, this means that there exists no vector $y'$ such that $y'\geq y$  and $y'_{i, s}>y_{i, s}$ for some student $i$ with type $t$. Now, since after this point the vector $y$ can only increase, we can never have a vector $y'$ that satisfies this property, and so students of type $t$ will always remain blocked.
\Halmos
\endproof

To show within-type envy-freeness, we use proof by contradiction. Assume there exists a student $i$ that envies another student $j$ of the same type. This implies that there exists some school $s$ such that:
$$\sum_{s':s'\succ_{i} s} x_{j, s'} > \sum_{s': s'\succ_{i} s} x_{i, s'}.$$

Let $A=\{s'\in M | s'\succ_i s \}$. We know that while it is possible for student $i$ to eat schools in $A$, she doesn't consume any schools in $M\setminus A$. Also, by Claim \ref{claim3}, we know that once she is blocked from eating some school in $A$, no other student of type $t$ can ever consume from that school. Since all the students consume schools at the same rate, this implies that $\sum_{s':s'\succ_{i} s} x_{j, s'} \leq \sum_{s': s'\succ_i s} x_{i, s'}$ for all $j\in N$ with type $t_i$, which is a contradiction.

To show ordinal efficiency, we use proof by contradiction. Let $x$ be the fractional assignment obtained by the eating algorithm. Consider another feasible assignment $y$ that stochastically dominates $x$. Note that the values of all the entries of $x$ have evolved over time, and  were initialized all to zero. Let $x_0$ denote the value of $x$ the last time $x$ was such that $x\leq y$. At that time, there must exist a student $i$ who is eating a school $s$ and $(x_0)_{i, s}=y_{i, s}$. At that point, student $i$ had the choice to eat any school $s'$ such that $(x_0)_{i, s'}<y_{i, s'}$, but she preferred to keep eating school $s$. Since $x_{i, s}>y_{i, s}$, $y_i$ cannot  stochastically dominate $x_i$ with respect to $\succ_i$,
 which is a contradiction.

Property (iii) is a direct implication of Lemma \ref{implementtheorem}.
\Halmos
\endproof

We note that \citet{katta2006solution} generalize  Probabilistic Serial to weak preferences without distributional constraints. Their mechanism resembles ours in the sense that it updates  ``menus" by resolving flow problems.

To illustrate GPS, consider the example in Section \ref{sec:ex}. Each student  first eats from her favorite school (at rate 1). After half a unit of time, each student has consumed half of her favorite school, and the  vector of consumed probability shares is no longer extendable. All students then switch to eat from their next favorite school. Again, after half a unit of time, each student has consumed half of their second favorite school and now each student has consumed one unit. Finally, the fractional assignment is  implemented  as a lottery over approximately feasible allocations.

\subsection{An impossibility result}

\cite{bogomolnaia2001new} find that no mechanism is ordinally efficient, envy-free, and strategyproof in the context of allocating indivisible objects to homogeneous agents. However, they show that the PS mechanism is weakly strategyproof. We prove that with distributional constraints, weak strategyproofness cannot hold when  both within-type envy-freeness and ordinal efficiency are required. (A similar impossibility for  the allocation problem without distributional constraints was established for the case in which preferences are not necessarily strict  \citep{katta2006solution}.)


\begin{theorem}
	In the school choice problem with distributional constraints, no mechanism  is ordinally efficient, within-type envy-free, and weakly strategyproof.
\end{theorem}
\proof{Proof.}
Consider three  schools $s_1, s_2, s_3$ and two students $i, j$ of type $t$. 
There are three additional types  $t_1,t_2,t_3$ other than $t$ and one student of each of those types. We have the following constraints for $s\in\{s_1,s_2\}$:
\begin{align*}
	1\leq x_{t_1, s}+x_{t_2, s} \leq 1,\\
	1\leq x_{t_2, s}+x_{t_3, s} \leq 1,\\
	1\leq x_{t_3, s}+x_{t_1, s} \leq 1,\\
	0\leq x_{t, s}+x_{t_1, s} \leq 1,
\end{align*}
where $x_{t, s}$ represents the number of students of type $t$ assigned to school $s$. Note that in the above inequalities, the values of $x_{t_1, s}, x_{t_2, s}$, and $x_{t_3, s}$ are all uniquely determined to be 0.5. Therefore, in each school $s_1$ and $s_2$, there is an upper bound of $0.5$ for type $t$. 
School $s_3$ imposes no constraints.

Now suppose that $i$ and $j$ have the following preferences: 
\begin{align*}
\succ_i:s_1 \succ_i s_2 \succ_i s_3,\\
\succ_j:s_2 \succ_j s_3 \succ_j s_1.
\end{align*}

Assume $j$ reports her true preferences and $i$ misreports $\succ'_i:s_2 \succ'_i s_1 \succ'_i s_3$. By ordinal efficiency, both $s_1$ and $s_2$ should be filled up to 0.5 with students $i$ and $j$. In addition, $s_1$ should be filled with student $i$.  Due to within-type envy-freeness,  $i$ and $j$ must each be assigned 0.25 of $s_2$. Therefore $i$'s assignment in schools $s_1, s_2, s_3$  must be $(0.5, 0.25, 0.25)$ and $j$'s assignment must be $(0, 0.25, 0.75)$.
So when $i$ and $j$   both report truthfully, $i$ must be assigned 0.5 of $s_1$ and at least 0.25 of $s_2$. This implies that $j$ is assigned at most 0.25 of $s_2$.  Denote the assignment of $j$ in this case by $x_j$.

Using this observation, we claim that $j$ will benefit from misreporting her preferences. Assume $i$ reports her true preferences and $j$ misreports $\succ'_j:s_2 \succ'_j s_1 \succ'_j s_3$. Denote the assignment under these reports by $x'$. By ordinal efficiency, $s_1$ and $s_2$ should be filled up to $0.5$. Moreover, $i$ and $j$ cannot both have positive assignment probabilities to $s_1$ and $s_2$ because they will benefit from exchanging  probability shares. In combination with  within-type envy-freeness, this implies that $j$'s assignment must be $(0, 0.5, 0.5)$.  This means that $x'_j \neq x_j$ and $x'_j (sd)_{\succ_j} x_j$, contradicting weak strategyproofness. 
\Halmos
\endproof



\section{Conclusion}
We studied the assignment problem under distributional constraints and privately known preferences. There may be numerous Pareto-efficient assignments,  which can vary significantly in the number of assigned students. The  mechanisms we introduced result in assignments that  may violate each lower- and upper-bound quota by at most the number of students' types, but  can assign  as many students as can be assigned via a fractional solution. While our generalization of the serial dictatorship is strategyproof,  we demonstrate that distributional constraints introduce a new barrier to achieving weak strategyproofness in combination with within-type envy-freeness and ordinal efficiency.


\ACKNOWLEDGMENT{Ashlagi gratefully acknowledges the research support of  National Science Foundation grant SES-1254768 and the support of  U.S-Israel  Binational Science Foundation grant 2012344.\\
Amin Saberi's research is supported by NSF award 1812919 and ONR Award N000141612893.

}

\bibliographystyle{informs2014}
\bibliography{referencesbib}

\begin{APPENDICES}
\section{Illustration of the Serial Dictatorship Algorithm with Dynamic Menus}
\label{app-example}

In Section \ref{sec:rsd} we illustrated the performance of Algorithm \ref{alg1} on a simple example. Here, we expand on that by analyzing a more detailed example.
Suppose there are two schools $s_1$ and $s_2$ and seven students $i_1, i_2, \ldots, i_7$ whose types are $t_1$, $t_2$, $t_3$, $t_4$, $t_5$, $t_1$, and $t_2$ respectively. Students $i_2$ and $i_4$ prefer $s_2$ over $s_1$ and the rest of the students prefer $s_1$ over $s_2$. We are given the following distributional constraints:

\begin{align*}
	&1\leq x_{t_1, s_1}+x_{t_2, s_1} \leq 1,&1\leq x_{t_3, s_2}+x_{t_4, s_2} \leq 1,\\
	&1\leq x_{t_2, s_1}+x_{t_3, s_1} \leq 1,&1\leq x_{t_4, s_2}+x_{t_5, s_2} \leq 1,\\
	&1\leq x_{t_3, s_1}+x_{t_1, s_1} \leq 1,&1\leq x_{t_5, s_2}+x_{t_3, s_2} \leq 1,\\
	&0\leq x_{t_1, s_1}+x_{t_2, s_1}+x_{t_3, s_1} \leq 2,&0\leq x_{t_1, s_2}+x_{t_2, s_2}+x_{t_3, s_2} \leq 2.
\end{align*}

Observe that given these quotas, in any feasible solution, each of types $t_1, t_2, $ and $t_3$ will be assigned to school $s_1$ with probability  $0.5$ and each of types $t_3, t_4, $ and $t_5$ will be assigned to school $s_2$ with probability  $0.5$ as well. Assume the order of students to be $i_1, i_2,\ldots,i_7$. In the first assignment step,  $i_1$ is partially  assigned to $s_1$ (with 0.5), and is added to the set of partially assigned students. In the first resolution step, no school is critical for type $t_1$.

In the second round, $i_2$ is completely assigned to $s_2$. This time, in the resolution step, $s_2$ is critical for type $t_1$ and $f(t_1, s_2)=0.5$. Therefore, $s_2$ is used to resolve the assignment of $i_1$ to $s_1$ by applying the $(i_1, s_2)$-updates. After this procedure, the constraints of schools $s_1$ and $s_2$ are updated as follows:

\begin{align*}
	&1.5\leq x_{t_1, s_1}+x_{t_2, s_1} \leq 1.5,&1\leq x_{t_3, s_2}+x_{t_4, s_2} \leq 1,\\
	&1\leq x_{t_2, s_1}+x_{t_3, s_1} \leq 1,&1\leq x_{t_4, s_2}+x_{t_5, s_2} \leq 1,\\
	&1.5\leq x_{t_3, s_1}+x_{t_1, s_1} \leq 1.5,&1\leq x_{t_5, s_2}+x_{t_3, s_2} \leq 1,\\
	&0.5\leq x_{t_1, s_1}+x_{t_2, s_1}+x_{t_3, s_1} \leq 2.5,&-0.5\leq x_{t_1, s_2}+x_{t_2, s_2}+x_{t_3, s_2} \leq 1.5.
\end{align*}

In the next assignment step, since $f(t_3, s_1)=0.5$, $i_3$ is partially assigned to $s_1$. In the following resolution step, $s_2$ is used to resolve her assignment and after the $(i_3, s_2)$-updates the constraints are updated to:

\begin{align*}
	&1.5\leq x_{t_1, s_1}+x_{t_2, s_1} \leq 1.5,&0.5\leq x_{t_3, s_2}+x_{t_4, s_2} \leq 0.5,\\
	&1.5\leq x_{t_2, s_1}+x_{t_3, s_1} \leq 1.5,&1\leq x_{t_4, s_2}+x_{t_5, s_2} \leq 1,\\
	&2\leq x_{t_3, s_1}+x_{t_1, s_1} \leq 2,&0.5\leq x_{t_5, s_2}+x_{t_3, s_2} \leq 0.5,\\
	&1\leq x_{t_1, s_1}+x_{t_2, s_1}+x_{t_3, s_1} \leq 3,&-1\leq x_{t_1, s_2}+x_{t_2, s_2}+x_{t_3, s_2} \leq 1.
\end{align*}

Next, we have $f(t_4, s_2)=0.5$, and therefore $i_4$ is partially assigned to $s_2$. In the resolution step $s_1$ is used to resolve this partial assignment. After the $(i_4, s_1)$-updates the constraints are:

\begin{align*}
	&1.5\leq x_{t_1, s_1}+x_{t_2, s_1} \leq 1.5,&1\leq x_{t_3, s_2}+x_{t_4, s_2} \leq 1,\\
	&1.5\leq x_{t_2, s_1}+x_{t_3, s_1} \leq 1.5,&1.5\leq x_{t_4, s_2}+x_{t_5, s_2} \leq 1.5,\\
	&2\leq x_{t_3, s_1}+x_{t_1, s_1} \leq 2,&0.5\leq x_{t_5, s_2}+x_{t_3, s_2} \leq 0.5,\\
	&1\leq x_{t_1, s_1}+x_{t_2, s_1}+x_{t_3, s_1} \leq 3,&-1\leq x_{t_1, s_2}+x_{t_2, s_2}+x_{t_3, s_2} \leq 1.
\end{align*}

In the next assignment step, we have $f(t_5, s_1)=0.5$ and $i_5$ is partially assigned to $s_1$. In the resolution step, $s_2$ is used to resolve this assignment. After the $(i_5, s_2)$ updates, the constraints are:

\begin{align*}
	&1.5\leq x_{t_1, s_1}+x_{t_2, s_1} \leq 1.5,&1\leq x_{t_3, s_2}+x_{t_4, s_2} \leq 1,\\
	&1.5\leq x_{t_2, s_1}+x_{t_3, s_1} \leq 1.5,&1\leq x_{t_4, s_2}+x_{t_5, s_2} \leq 1,\\
	&2\leq x_{t_3, s_1}+x_{t_1, s_1} \leq 2,&0\leq x_{t_5, s_2}+x_{t_3, s_2} \leq 0,\\
	&1\leq x_{t_1, s_1}+x_{t_2, s_1}+x_{t_3, s_1} \leq 3,&-1\leq x_{t_1, s_2}+x_{t_2, s_2}+x_{t_3, s_2} \leq 1.
\end{align*}

Next, we have $f(t_1, s_1)=0$, and $f(t_1, s_2)=0$ and therefore $i_6$ is assigned to the outside option. Finally, when $i_7$ is about to get assigned, we have $f(t_2, s_1)=0.5$ and therefore $i_7$ is partially assigned to $s_1$. In the following resolution step, we use the outside option, i.e. $\phi$ to resolve this partial assignment. After the $(i_7, \phi)$-updates, the constraints are:

\begin{align*}
	&2\leq x_{t_1, s_1}+x_{t_2, s_1} \leq 2,&1\leq x_{t_3, s_2}+x_{t_4, s_2} \leq 1,\\
	&2\leq x_{t_2, s_1}+x_{t_3, s_1} \leq 2,&1\leq x_{t_4, s_2}+x_{t_5, s_2} \leq 1,\\
	&2\leq x_{t_3, s_1}+x_{t_1, s_1} \leq 2,&0\leq x_{t_5, s_2}+x_{t_3, s_2} \leq 0,\\
	&1.5 \leq x_{t_1, s_1}+x_{t_2, s_1}+x_{t_3, s_1} \leq 3.5,&-1\leq x_{t_1, s_2}+x_{t_2, s_2}+x_{t_3, s_2} \leq 1.
\end{align*}

\end{APPENDICES}




\end{document}